\documentclass[11pt]{article}
\usepackage{amsthm,amssymb,amsfonts,amsmath,amscd}
\usepackage{nccmath}
\usepackage{graphicx}
\usepackage[pdfa]{hyperref}

\makeatletter \@addtoreset{equation}{section} \makeatother

\newtheorem{theorem}{Theorem}[section]
\newtheorem{lemma}{Lemma}[section]

\newtheorem{proposition}{Proposition}[section]

\newcommand{\mdet}{\mathrm{det}}
\newcommand{\intd}{\displaystyle\int}
\newcommand{\Tr}{\mathrm{Tr}\,}
\newcommand{\Str}{\mathrm{Str}\,}
\newcommand{\Sdet}{\mathrm{Sdet}\,}
\newcommand{\E}{\mathbb{E}}
\newcommand{\p}{\mathbb{P}}

\begin{document}
\title{The least singular value of the general deformed Ginibre ensemble}
\author{ Mariya Shcherbina
\thanks{Institute for Low Temperature Physics, Kharkiv, Ukraine, e-mail: shcherbi@ilt.kharkov.ua} \and
 Tatyana Shcherbina
\thanks{ Department of Mathematics, University of Wisconsin - Madison, USA, e-mail: tshcherbyna@wisc.edu. }}
\date{}
\maketitle

\begin{abstract}
We study the least singular value of the $n\times n$ matrix $H-z$ with $H=A_0+H_0$, where $H_0$ is drawn from the complex Ginibre ensemble of matrices with iid Gaussian entries, and $A_0$ is some general $n\times n$ matrix with complex entries
(it can be random and in this case it is independent of $H_0$). Assuming some rather general assumptions on $A_0$, we prove an
optimal tail estimate on the least singular value in the regime where $z$ is around the spectral edge of $H$ thus generalize the recent result
of Cipolloni, Erd\H{o}s, Schr\"{o}der \cite{CiErS:20} to the case $A_0\ne 0$. The result improves the classical bound by Sankar, Spielman and Teng \cite{SST:06}.

\end{abstract}

\section{Introduction}\label{s:1}
Consider random  $n\times n$ matrices
\begin{equation}\label{H_1}
H=A+H_0,
\end{equation}
where $A$ is some general $n\times n$ matrix with complex entries, and $H_0$ is drawn from the complex Ginibre ensemble,
i.e. $H_0$ has i.i.d. complex Gaussian entries $\{h_{ij}^{(0)}\}_{i,j=1}^n$ such that
\[
\E[h_{ij}^{(0)}]=0, \quad \E[|h_{ij}^{(0)}|^2]=1/n,\quad \E[(h_{ij}^{(0)})^2]=0.
\]
Deformation $A$ can be deterministic or random (but in this case it is independent of $H_0$).

Such matrices are important in communication theory, where $A$ is considered as a \textit{signal}, and
$H_0$ as a \textit{noise} matrix. In particular, one is interested in effective numerical solvability of a large
system of linear equations $H x=b$ which is determined by the behaviour of the smallest singular value
$\sigma_1(H)$ of $H$.

The classical bound of Sankar, Spielman and Teng \cite{SST:06} states that the smallest singular value $\sigma_1(H)$ is of order not smaller than $n^{-1}$
(equivalently, the smallest eigenvalue $\lambda_1(HH^*)$ of $HH^*$ is of order not smaller that $n^{-2}$), i.e.
\begin{equation}\label{SST}
\p\Big(\lambda_1(HH^*)=(\sigma_1(H))^2\le x/n^2\Big)\lesssim x,\quad x>0.
\end{equation}
up to logarithmic corrections, uniformly in $A$. If $H_0$ is a real Ginibre matrix, then
the bound is $\sqrt{x}$ instead of $x$.

The bound is proved to be optimal for the case of pure complex Ginibre ensemble (i.e. $A=0$), see \cite{Ed:88}.
Similar results for matrices with iid entries (i.e. $A=0$, but $H_0$ is not necessary Gaussian) are also known
(see, e.g., \cite{RV:08}, \cite{TV:10} and references therein).

The matrix $HH^*$ can be considered as the so-called deformed Laguerre ensemble, and its limiting eigenvalue distribution is well-known, see \cite{DSil:07}.
Moreover, according to \cite{BaSil:12}, almost surely no eigenvalues lie outside  any finite support neighbourhood  of the
limiting measure. Therefore, we can have three possible situations: $0$ is away of the support of limiting spectral measure, and then $\lambda_1(HH^*)$ has a
positive constant lower bound; $0$ lies in the bulk of the spectrum, and then we get (\ref{SST});  the intermediate regime when $0$ is near the edge of the
spectrum.

In the bulk regime the lower bounds on $\lambda_1(HH^*)$ with quite general $A$ and even with the non-Gaussian $H_0$ (with iid elements) have been obtained
in \cite{TV:07},\cite{TV:10_1} (but not uniformly in $A$); see also \cite{C:18}, \cite{T:20} beyond the i.i.d. case.

The edge regime is much less studied. However, as it was shown in \cite{CiErS:20} for the case of the constant diagonal shift of the Ginibre ensemble, i.e. $A=-z I$, 
the bound (\ref{SST}) can be improved in the edge regime $|z|\sim 1$.
The aim of the current paper is to extend the result of \cite{CiErS:20} to the case $A=A_0-z I$ with a rather general complex deformation $A_0$.

Another important source of motivation  is that an effective lower tail bound on the least singular
value of $H-z$ is an essential ingredient for the study of eigenvalues distribution of large non-Hermitian matrices.
In particular, the results of Cipolloni, Erd\H{o}s, Schr\"{o}der \cite{CiErS:20} was used in their subsequent work \cite{CiErS:ed}  to remove the four moment matching condition in the classical edge universality result for non-Hermitian random matrices with iid entries  by Tao, Vu \cite{TV:15}.
Better understanding of higher order correlation functions of the shifted Ginibre ensemble in the bulk is expected to  help to do the same thing for the
universality in the bulk. In addition, \cite{CiErS:20} is also an important ingredient for the recent CLT result \cite{CiErS:CLT}.

To obtain the lower bound on the least singular value of $H-z=A_0+H_0-z$, we are going to study  the resolvent of 
\begin{equation}\label{Y}
Y(z)=(H-z)(H-z)^*.
\end{equation}
More precisely, we are going to obtain the integral representation and the precise asymptotic behaviour of
\begin{equation}\label{T}
T(z,\varepsilon)=\E\Big\{n^{-1}\Tr\,(Y(z)+\varepsilon^2)^{-1}\Big\},
\end{equation}
which is related to the average density of states (or one-point correlation function) of $Y(z)$ at $0$.
For the small $\varepsilon$ the main contribution to (\ref{T}) comes from the lowest eigenvalue $\lambda_1(Y(z))$, hence 
an upper estimate on (\ref{T}) implies a lower tail bound on $\lambda_1(Y(z))$.

The ensemble $Y(z)$ was extensively studied in \cite{BenP:05} by using so-called Br\'{e}zin-Hikami formulas (see \cite{BrHi:98}).
 Theorem 7.1 of \cite{BenP:05} allows to represent even higher order correlation function of $Y(z)$ as the determinants of a certain kernel which 
can be computed as a double integral involving the Bessel kernel. Paper \cite{BenP:05} 
did not analyse the resulting one point function (and, generally, used this integral representation to study the case when $H=A_0+H_0$ is Gaussian divisible ensemble only), but one can use this integral representation to analyze $T(z,\varepsilon)$.

In this paper, however, we use a different approach because of two main reasons.
First, there is no analogue of Br\'{e}zin-Hikami formulas for the real symmetric case, and, second, the local 
study of eigenvalues distribution of  $H$ requires the analysis of  the joint distribution 
the smallest singular values of $H-z_1$, $H-z_2$,\ldots, $H-z_k$ for different $z_i$ which is not covered by 
the determinantal formulas of  \cite{BenP:05}. 

Our approach is based on supersymmetry (SUSY) techniques and is expected to be more robust, in particular, it is available in the case of real symmetry. It based on the fact
that SUSY  techniques allows to rewrite the main spectral characteristics of random matrices
(such as density of states, correlation functions, etc.) as an integral containing both complex and
Grassmann (anticommuting) variables. The method is widely used in
the physics literature (see, e.g., \cite{Ef},\cite{M:00}), but the rigorous analysis of such integral representations usually is quite difficult.
However, the method was successfully applied to the rigorous study of some random matrix ensembles, including the most
successful applications to the Gaussian random band matrices (see \cite{EB:15}, \cite{DL:16}, \cite{DPS:02}, \cite{SS:den} -- \cite{TS:real}), as well as
 to the study of overlaps of non-Hermitian Ginibre eigenvectors \cite{Fyo:18}.

The method of \cite{CiErS:20} for the case of $A_0=0$ also utilizes SUSY. Their  technique is based on the superbosonization formula by Littelmann, Sommers and Zirnbauer
\cite{SupB:08} which significantly reduces the number integration variables: instead of $2n$ real and $2n$ Grassmann variables of integration
one can get the integral over the two complex variables only.
 However, this formula is not applicable in the case of general $A_0$, so we need to use another approach based on the Hubbard-Stratonovich transformation (see (\ref{Hub_C}) -- (\ref{Hub_Gr}) below)
 and Fourier transform (see Section 2 for the discussion).

Now let us formulate the main results of the paper. Consider the matrix
\begin{align*}
H(z)=A_0+H_0-z,
\end{align*}
where $H_0$ is a complex Ginibre matrix defined in (\ref{H_1}), and $A_0$ is some general $n\times n$ matrix with complex entries (which can be random, and
in this case is independent of $H_0$) satisfying the following conditions

\textit{Assumptions (A1)-(A4)}:
\begin{enumerate}
\item[(A1)] There are some $M,d>0$ such that
\begin{align*}
\mathrm{Prob}\Big\{n^{-1}\sum_{i,j=1}^n|A_{0,ij}|^2<M\Big\}\ge 1- n^{-1-d}.
\end{align*}
\item[(A2)] 
For almost all $z$ normalized counting measure of eigenvalues  of the matrix\\ $Y_0(z):=(A_0-z)(A_0-z)^*$ converges, as $n\to\infty$, to some limiting
measure $\nu_z$;
\item[(A3)] Denote $\sigma_0=\{z:0\in \mathrm{supp}\,\nu_z\}$,  $\sigma_{\epsilon}$ - $\epsilon$-neighbourhood of $\sigma_0$,
and $s_n(A_0)$ -  spectrum of $A_0$.  Set 
\begin{align*}
&\Omega^{(1)}_\epsilon=\{\omega: s_n(A_0)\subset\sigma_{\epsilon}\},
\\ \notag
&\Omega^{(2)}_\epsilon=\{\omega: \sup_{z\not\in\sigma_{\epsilon}}\Big|n^{-1}\Tr Y_0^{-1}(z)-\int \lambda^{-1} d\nu_z(\lambda)\Big|\le C_\epsilon n^{-1/2-d_0}\},
\end{align*}
where $d_0>0$ is some fixed number.

Then there is some  $d>0$  such that
\begin{align*}
&\mathrm{Prob}\{\Omega_\epsilon^{(1)}\cap \Omega_\epsilon^{(2)}\}\ge 1- Cn^{-1-d};
\end{align*}
\item[(A4)] There is $d_1>0$ such that for some sufficiently small $\epsilon_0,\varepsilon$ 
if we set
\begin{align*}  
\Omega^{(3)}=\Big\{\omega:\inf_{z\in\sigma_{\epsilon_0}}n^{-1}\Tr\,\big(Y_0(z)+\varepsilon^2\big)^{-1}>1+d_1\Big\},
\end{align*}
then
\begin{align*}  
\mathrm{Prob}\{\Omega^{(3)}\}>1-C'_\epsilon n^{-1-d}
\end{align*}

\end{enumerate}

Below we give a few examples of $A_0$ satisfying the assumption (A1)-(A4):
\begin{enumerate}
\item[(1)]   $A^{(n)}$ is a sequence of matrices  whose limiting spectrum consists of  a finite number of points
$\sigma_0 =\{\zeta_1,\dots,\zeta_k\}$  and we have some large deviation
type bound for  distribution of $\{\zeta_j^{(n)}\}$.

\item[(2)]  $A_0=A^*_0$ - any classical  hermitian model like Wigner matrices, sample covariance matrices, sparse matrices,
etc.

\item[(3)]   $A_0$ is a diagonal matrix $A_0=\mathrm{diag}\{\zeta_j^{(n)}\}$
with $\{\zeta_j^{(n)}\}$ having limiting distribution  with a compact finitely connected support $\sigma_0$ 
with a smooth boundary and a non zero limiting density and such that we have some large deviation type bound
(like (A3), (A4)).

\item[(4)]  $A_0$ is a Ginibre matrix with i.i.d. entries having all moments. Then, according to \cite{TVKr:10}, its 
normalized counting eigenvalue distribution  converges to a circle low  and we have a large deviation type bound (A3), (A4)
due to the result of \cite{AEK:20}.

\end{enumerate}
According to the result of \cite{TVKr:10}, under assumption A1-A2 there exists a non-random probability measure $\mu$ on the complex plain
which is a limit of the normalized counting measure of eigenvalues of $H$.   In addition, for almost all $z\in\mathbb{C}$ there exists
a non-random probability measure $\eta_z$ on $\mathbb{R}$ which is a limit of the normalized counting measure of eigenvalues of $Y(z)$
defined in (\ref{Y}) (see \cite{DSil:07}). The support $D$ of the limiting measure $\mu$ is not so easy to describe, however, 
according to \cite{BoCap:14}, under an additional assumption 
\[(A')\quad D=\mathrm{supp}\,\mu=\{z: 0\in\mathrm{supp}\,\eta_z\},
\]
it takes the nice form:
 \begin{align}\label{dD}
  D=\{z:\limsup_{\varepsilon\to 0}n^{-1}\Tr\,\big(Y_0(z)+\varepsilon^2\big)^{-1}\ge 1\}.
 \end{align}
Notice that the result of \cite{BoCap:14} also  includes $\sigma_0$ of (A3) to $\mathrm{supp}\,\mu$, but the assumptions  (A3) -- (A4)  guarantee  that $\sigma_0\in D$  with probability 1, so $\mathrm{supp}\,\mu$ coincides with (\ref{dD}). Remark also that the authors of \cite{BoCap:14} mentioned that they are not aware of any examples 
where $(A')$ fails to hold.
 
Assumption (A4)  guarantees that there is $\epsilon>0$ such that the boundary $\partial D\cap\sigma_\epsilon=\emptyset$.
 Thus,  $\partial D$ is a level line of the smooth function 
 \begin{align}\label{cal-F}
 \mathcal{F}(z):=\int \lambda^{-1} d\nu_z(\lambda),
 \end{align}
and hence $\partial D$ is a set of  piece-wise smooth closed curves enclosing $\sigma_0$ of (A3). 
  
  
The first theorem gives an asymptotic behavior of $T(z,\varepsilon)$ in the edge regime where $z$ lies inside the spectrum of $A_0+H_0$, but the distance to the boundary of the spectrum is of order $n^{-1/2}$.
 \begin{theorem}\label{t:1} 
 Given $z\in D$ such that $dist\{z,\partial D\}=\tilde\delta^2 n^{-1/2}$ and $\partial D$ is a smooth curve in some neighbourhood
of $z$,  and   given $\varepsilon^2=\tilde\varepsilon^2 n^{-3/2}$
with $\tilde\delta, \tilde\varepsilon\sim 1$,   under assumptions (A1)-(A4) and (A') we have: 
\begin{align}\label{t1.1}
T(z,\varepsilon)=n^{1/2}I(\tilde\delta,\tilde\varepsilon),
\end{align}
where $I(\tilde\delta,\tilde\varepsilon)\sim 1$ is represented by some integral, depending on $\tilde\delta,\tilde\varepsilon$.
 \end{theorem}
The  next theorem gives an asymptotic behavior of $T(z,\varepsilon)$ in the bulk regime where $z$ is well inside the spectrum of $A_0+H_0$. In this regime the bound (\ref{SST})  becomes optimal.
  \begin{theorem}\label{t:2} 
 Under assumptions (A1)-(A4) and (A'), if we choose $z\in D$, $dist\{z,\partial D\}>\delta^2\sim 1 $ and  $\varepsilon=(\delta n)^{-1}\tilde\varepsilon $
with $ \tilde\varepsilon\sim 1$,  then
\begin{align}\label{t2.1}
T(z,\varepsilon)=n\tilde I,
\end{align}
where $\tilde I(\tilde\varepsilon)\sim 1$ is represented by some integral, depending on $\tilde\varepsilon$.
 \end{theorem}
 The transition regime between Theorem \ref{t:1} and Theorem \ref{t:2} is given by
  \begin{theorem}\label{t:3} 
 Given $z\in D$ such that $dist\{z,\partial D\}=\delta^2 $ with $1\gg\delta^2\gg n^{-1/2}$ and $\partial D$ is a smooth curve in some neighbourhood
of $z$, and given $\varepsilon=(\delta n)^{-1}\tilde\varepsilon $
with $ \tilde\varepsilon\sim 1$,  under assumptions (A1)-(A4) and (A') we have:
\begin{align}\label{t3.1}
T(z,\varepsilon)=(\delta/\varepsilon)\tilde I (\tilde\varepsilon)(1+O((\delta^4n)^{-1/2})),
\end{align}
where $\tilde I(\tilde\varepsilon)\sim 1$ is represented by some integral, depending on $\tilde\varepsilon$.
 \end{theorem}
Since the main contribution to (\ref{T}) comes from the lowest eigenvalue $\lambda_1(Y(z))$, by a straightforward Markov inequality Theorems \ref{t:1} -- \ref{t:3} give the following corollary
\begin{theorem}\label{t:tail}
Under assumptions (A1)-(A4) and (A'), uniformly in $z\in D$,  $dist\{z,\partial D\}\ge C n^{-1/2}$  and $0<x<C$ 
\begin{equation}\label{tail_b}
 \p\Big(\lambda_1(Y(z))\le c(n,z) x\Big)\lesssim (1+|\log x| ) x,
\end{equation}
where
\[
c(n,z)=\min\Big\{\frac{1}{n^{3/2}}, \frac{1}{n^2\cdot \mathrm{dist}\{z,\partial D\}}\Big\}
\]
\end{theorem}
Notice that conditions (A3) and (A4) allow us to prove Theorems \ref{t:1} -\ref{t:3}  only for \\
$\omega\in \Omega_{\epsilon_0/2}^{(1)}\cap \Omega_{\epsilon_0/2}^{(2)}\cap \Omega^{(3)}$ since 
on the complement we can use the trivial bound
\begin{align*}
n^{-1}\Tr\,(Y(z)+\varepsilon^2)^{-1}\le \varepsilon^{-2}
\end{align*}
combined with the inequality
\begin{align*}
\mathrm{Prob}\{\big(\Omega_{\epsilon_0/2}^{(1)}\cap \Omega_{\epsilon_0/2}^{(2)}\cap \Omega^{(3)}\big)^c\}\le Cn^{-1-d}.
\end{align*}
These bounds give us that the  contributions of $\big(\Omega_{\epsilon_0/2}^{(1)}\cap \Omega_{\epsilon_0/2}^{(2)}\cap \Omega^{(3)}\big)^c$ are small  in comparison with the main terms of the r.h.s. of (\ref{t1.1}) -- (\ref{t3.1}).

The paper is organized as follows.  In Section 2 we obtain the SUSY integral representation of $T(\varepsilon,z)$ of (\ref{T}). 
Section 3 -- 5 deal with the proof of Theorems \ref{t:1} -- \ref{t:3} respectively. The brief outline of SUSY techniques is given in Appendix.

\section{Integral representation of $T(z,\varepsilon)$}
The aim of this section is to derive an integral representation for $T(z,\varepsilon)$.
 \begin{proposition}\label{p:repr} 
Given (\ref{T}), we have
 \begin{align}\label{p1.1}
T(z,\varepsilon)=&\frac{n^3}{2\pi^3\varepsilon}\int_{-\infty}^\infty (u_1+\varepsilon)du_1 d u_2  \int_{L}dt_1dt_2 \int_{0}^\infty r_1r_2dr_1dr_2
\\&\varphi(u_1,  u_2,  t_1, t_2)
\exp\{nF_1(u_1,u_2)\}\exp\{nF_2(t_1, t_2, r_1, r_2)\}\notag
\end{align}
where $L:=\mathbb{R}+i\varepsilon_0$
\begin{align}\label{p1.2}
F_1(u_1,u_2)&=\mathcal{L}_n(u_1^2+u_2^2)-(u_1+\varepsilon)^2-u_2^2\\
F_2(t_1, t_2, r_1, r_2)&=-\mathcal{L}_n(-t_1t_2)-(r_1r_2)^2-ir_1r_2(t_1+t_2)
-\varepsilon(r_1^2+r_2^2)\notag\\
\varphi(u_1,  u_2,  t_1, t_2)=&
\Big(1-\frac{1}{n}\Tr G(-t_1t_2)+\frac{|u|^2}{n}\Tr G(u\bar u)G(-t_1t_2)\Big)^2\notag\\
&+t_1t_2|u|^2\Big(\frac{1}{n}\Tr G(u\bar u)G(-t_1t_2)\Big)^2\notag\\
&-\frac{|u|^2+t_1t_2}{n^2}\Big(\Tr G(u\bar u)G^2(-t_1t_2)-|u|^2\Tr G^2(u\bar u)G^2(-t_1t_2)\Big)\notag
\end{align}
 with 
  \begin{align}\label{L,G}
\mathcal{L}_{n}(x):=& n^{-1}\log\det(Y_0(z)+x),\quad Y_0(z):=(A_0-z)(A_0-z)^*\\
 G(x):=&(Y_0(z)+x)^{-1}.\notag
 \end{align}

 \end{proposition}

\noindent{\it Proof of Proposition \ref{p:repr}}

The first step is to rewrite $T(z,\varepsilon)$ in the following way:
\[T(z,\varepsilon)=\frac{1}{2\varepsilon n}\frac{d}{d\varepsilon_1}\mathcal{Z}(\varepsilon,\varepsilon_1)\Big|_{\varepsilon_1=\varepsilon},\quad
\mathcal{Z}(\varepsilon,\varepsilon_1)=\E\Big\{\frac{\det(Y(z)+\varepsilon_1^2)}{\det(Y(z)+\varepsilon^2)} \Big\}.\]
To obtain the integral representation of $\mathcal{Z}(\varepsilon,\varepsilon_1)$ we use the supersymmetric approach (SUSY).  
For the reader's convenience  a brief outline of the techniques is given
in Appendix.

Introduce Grassmann variables and complex variables
\begin{align}\label{psi}
&\Psi_l=(\psi_{l1},\dots,\psi_{ln})^t,\quad \bar\Psi_1=(\bar\psi_{l1},\dots,\bar\psi_{ln})^t,\quad l=1,2, \quad -\quad \hbox{Grassmann};\\
&X_l=(x_{l1},\dots,x_{ln})^t,\quad \bar X_l=(\bar x_{l1},\dots,\bar x_{ln})^t,\quad l=1,2,\quad -\quad \hbox{complex}.\notag
\end{align}
We are going to use the standard  linearisation  formula:
\begin{align*}
\det(Y(z)+\varepsilon^2)=\det \tilde Y(z,\varepsilon), \quad \tilde Y(z,\varepsilon)
=\left(\begin{array}{cc}-\varepsilon&i(H-z)\\i(H-z)^*&-\varepsilon\end{array}\right).
\end{align*}
Since all eigenvalues of $\tilde Y(z,\varepsilon)$ have real part $-\varepsilon$, we can apply (\ref{G_C}), (\ref{G_Gr}) to write
\begin{align*}
&(\det(Y(z)+\varepsilon^2))^{-1}=\int d\bar X dX\exp\{ (\tilde Y(z,\varepsilon)X,\bar X)\},\\
&\det(Y(z)+\varepsilon_1^2)=\int d\bar \Psi d\Psi \exp\{ (\tilde Y(z,\varepsilon_1)\Psi,\bar \Psi)\},
\end{align*}
where we denoted $X=\left(\begin{matrix}X_1\\ X_2\end{matrix}\right)$, $\Psi=\left(\begin{matrix}\Psi_1\\\Psi_2\end{matrix}\right)$, and
\begin{align*}
d\bar XdX=\prod_{j=1}^n \dfrac{d\bar x_{1j} dx_{1j}d\bar x_{2j} dx_{2j}}{\pi^2},\quad d\bar \Psi d\Psi =\prod_{j=1}^n d\bar \psi_{1j} d\psi_{1j}d\bar \psi_{2j} d\psi_{2j}.
\end{align*}
Hence, we obtain
\begin{align}\label{Z_rep}
\mathcal{Z}(\varepsilon,\varepsilon_1)=&\int d\bar X dX d\bar \Psi d\Psi \,\, Z_{\varepsilon}(X) Z_{\varepsilon_1}(\Psi) E(\Psi,X),
\end{align}
where
\begin{align}\label{ZZE}
&Z_{\varepsilon}(X)=\exp\Big\{-\varepsilon((X_1,\bar X_1)+(X_2,\bar X_2))+i((A_0-z)X_1,\bar X_2)+i((A_0-z)^*X_2,\bar X_1)\},\\ \notag
&Z_{\varepsilon_1}(\Psi)=\exp\Big\{-\varepsilon_1((\Psi_1,\bar \Psi_1)+(\Psi_2,\bar \Psi_2))
+i((A_0-z)\Psi_1,\bar \Psi_2)+i((A_0-z)^*\Psi_2,\bar \Psi_1)\Big\},\\ \notag
&E(\Psi,X)= \E\Big\{\exp\Big\{i\sum_{i,j}(H_{ij}(x_{2j}\bar x_{1i}+\psi_{2j}\bar \psi_{1i})+\bar H_{ij}(x_{1i}\bar x_{2j} +\psi_{1i}\bar \psi_{2j})\Big\}\Big\}.
\end{align}
Taking the expectation with respect to $H_{ij}$ we get
\begin{align} \notag
E(\Psi,X)=&\exp\Big\{-\frac{1}{n}\sum_{i,j}(x_{2j}\bar x_{1i}+\psi_{2j}\bar \psi_{1i})(x_{1i}\bar x_{2j} +\psi_{1i}\bar \psi_{2j})\Big\}
\\ \notag
=&\exp\Big\{-\frac{1}{n}(X_1,\bar X_1)(X_2,\bar X_2)+\frac{1}{n}(\Psi_1,\bar \Psi_1)(\Psi_2,\bar \Psi_2)\\ \notag &
-\frac{1}{n}(\bar X_1,\Psi_1)(X_2,\bar \Psi_2)
-\frac{1}{n}(\bar X_2,\Psi_2)(X_1,\bar \Psi_1)\Big\}.
\end{align}
Notice that, in contrast to the case $A_0=0$ considered in \cite{CiErS:20},  for general $A_0$ the functions $Z_{\varepsilon}(X)$, $Z_{\varepsilon_1}(\Psi)$ of (\ref{ZZE})
are not the functions of variables $p_{i,j}=(X_i,\bar X_j)$, $q_{i,j}=(\Psi_i,\bar \Psi_j)$ only, so we cannot apply the superbosonization formula of \cite{SupB:08}.
Instead, we use the Hubbard- Stratonovich transformation (\ref{Hub_C}) to get
\begin{align}\label{Hub1}
e^{(\Psi_1,\bar \Psi_1)(\Psi_2,\bar \Psi_2)/n}=&\frac{n}{\pi}\int  \exp\{u'(\Psi_1,\bar \Psi_1)+\bar u'(\Psi_2,\bar \Psi_2) 
-nu'\bar u'\}du'_1d u'_2,\quad u'=u_1'+iu_2'.
\end{align}
Unfortunately, one cannot  use the same transformation for $e^{-(X_1,\bar X_1)(X_2,\bar X_2)/n}$ since the integral with respect to $X_1,X_2$ becomes divergent.
Instead, we can apply the Fourier transform formula. 

\noindent More precisely, set
\begin{align*}
r_1r_2&\Phi(r_1^2,r_2^2)\\=&\int_{(X_1,\bar X_1)=r_1^2}dX_1d\bar X_1\int_{(X_2,\bar X_2)=r_2^2}dX_2d\bar X_2\exp\{i((A_0-z)^*X_1,\bar X_2)+i((A_0-z)X_2,\bar X_1)\\&-\varepsilon(X_1,\bar X_1)-\varepsilon(X_2,\bar X_2)
-\frac{1}{n}(\bar X_1,\Psi_1)(X_2,\bar \Psi_2)
-\frac{1}{n}(\bar X_2,\Psi_2)(X_1,\bar \Psi_1)\}.
\end{align*}
We remark that $\Phi$ evidently depends also on $\Psi_1,\bar\Psi_1,\Psi_2,\bar\Psi_2$, but we omit these arguments here  in order to
simplify formulas. Changing $r_1\to nr_1$, $r_2\to nr_2$, we obtain from (\ref{Z_rep}) and (\ref{Hub1})
\begin{align}\label{Z_Phi}
&Z(\varepsilon,\varepsilon_1)=\dfrac{n^3}{\pi}\int_0^\infty\int r_1r_2\,\Phi(n r_1^2, n r_2^2) \cdot \exp\{-n r_1^2r_2^2\}\cdot Z_{\varepsilon_1}(\Psi)\\ \notag
&\times \exp\{u'(\Psi_1,\bar \Psi_1)+\bar u'(\Psi_2,\bar \Psi_2) 
-nu'\bar u'\} d\bar u'd u' d\Bar\Psi d\Psi dr_1dr_2.
\end{align}
Using the inverse Fourier transform formula we get
\begin{align*}
\Phi(nr_1^2,nr_2^2)=\frac{1}{(2\pi)^2}\int \hat \Phi(t_1,t_2)\, e^{-int_1r_1^2-int_2r_2^2} \,dt_1dt_2 ,
\end{align*}
where
\begin{align*}
 \hat \Phi(t_1,t_2)=&4\int e^{it_1r_1^2+it_2r_2^2}r_1r_2\,\Phi(r_1^2,r_2^2) \,dr_1dr_2 \\
 =&4\int_{\mathbb{C}^{2n}} d\bar X dX\,\exp\{(it_1-\varepsilon)(X_1,\bar X_1)+(it_2-\varepsilon)(X_1,\bar X_1)\\
 &+i((A_0-z)^*X_1,\bar X_2)+i((A_0-z)X_2,\bar X_1)\\
&-\frac{1}{n}(\bar X_1,\Psi_1)(X_2,\bar \Psi_2)
-\frac{1}{n}(\bar X_2,\Psi_2)(X_1,\bar \Psi_1)\}.
\end{align*}
Thus, we obtain
\begin{align*}
\Phi(nr_1^2,nr_2^2)=&\frac{1}{\pi^2}\int dt_1dt_2 \int d\bar X d X\,\exp\{i((A_0-z)^*X_1,\bar X_2)+i((A_0-z)X_2,\bar X_1)\\
&+it_1((X_1,\bar X_1)-nr_1^2)+it_2((X_2,\bar X_2)-nr_2^2)-\varepsilon(X_1,\bar X_1)-\varepsilon(X_2,\bar X_2)\\
&-\frac{1}{n}(\bar X_1,\Psi_1)(X_2,\bar \Psi_2)
-\frac{1}{n}(\bar X_2,\Psi_2)(X_1,\bar \Psi_1)\}.
\end{align*}
Let us make the change of variables 
\begin{align*}
X_{1}=&r_1X_1',\quad X_{2}=r_2X_2',\quad t_1=\frac{r_2}{r_1}t_1',\quad 
t_2=\frac{r_1}{r_2}t_2'
\end{align*}
  and denote
\begin{align*}
R=r_1r_2.
\end{align*}
Then we get 
\begin{align}\notag
\Phi(nr_1^2,nr_2^2)=&\frac{R^{2n}}{\pi^2}\int dt_1'dt_2' \int d\bar X'd X' \exp\{iR((A_0-z)^*X_1',\bar X_2')+iR((A_0-z)X_2',\bar X_1')\\ \notag
&+it_1'R((X_1',\bar X_1')-n)+it_2'R((X_2',\bar X_2')-n)-R(\varepsilon\frac{r_1}{r_2} (X_1',\bar X_1')+\varepsilon \frac{r_2}{r_1} (X_2',\bar X_2'))\\ \label{Phi_rew}
&-\frac{R}{n}(\bar X_1',\Psi_1)(X_2',\bar \Psi_2)
-\frac{R}{n}(\bar X_2',\Psi_2)(X_1',\bar \Psi_1)\}.
\end{align}
Now we use the Hubbard-Stratonovich transformation (\ref{Hub_Gr}) for the Grassmann variables:
\begin{align}\label{Hub2}
 e^{-R(\bar X_1,\Psi_1)(X_2,\bar \Psi_2)/n}=&\int d\eta_1 d\chi_1 \exp\{\chi_1(\bar X_1,\Psi_1)(R/n)^{1/2}+\eta_1(X_2,\bar \Psi_2)(R/n)^{1/2}+\chi_1 \eta_1\},\\
  e^{-R(\bar X_2,\Psi_2)(X_1,\bar \Psi_1)/n}=&\int  d\eta_2 d\chi_2\exp\{\chi_2(\bar X_2,\Psi_2)(R/n)^{1/2}+\eta_2(X_1,\bar \Psi_1)(R/n)^{1/2}+\chi_2 \eta_2\}.
\notag\end{align}
Substituting this and (\ref{Phi_rew}) to (\ref{Z_Phi}) we obtain finally
\begin{align}\notag
\mathcal{Z}&(\varepsilon,\varepsilon_1)=\frac{n^3}{\pi^{3}}\int_0^\infty\, dr_1dr_2\, R^{2n+1}\int dt_1'dt_2' du_1'du_2' \int d\bar X' dX' d\bar \Psi d\Psi \\
&\times\exp\Big\{iR((A_0-z)^*X_1',\bar X_2')+iR((A_0-z)X_2',\bar X_1')-nR^2\notag\\
&+it_1'R((X_1,\bar X_1)-n)+it_2'R((X_2,\bar X_2)-n)-R(\varepsilon\frac{r_1}{r_2} (X_1',\bar X_1')+\varepsilon \frac{r_2}{r_1} (X_2',\bar X_2'))\Big\}\notag
\end{align}
\begin{align}
&\times P(X,\Psi,R) \cdot \exp\Big\{u'(\Psi_1,\bar \Psi_1)+\bar u'(\Psi_2,\bar \Psi_2)-n|u'|^2\}\cdot Z_{\varepsilon_1}(\Psi)\notag\\
=&\frac{n^3}{\pi^{3}} \int_0^\infty dr_1dr_2\, R^{2n+1}e^{-nR^2}e^{-n|u'|^2}\int dt_1' dt_2' du_1' du_2' d\eta d\chi \,\, e^{\chi_1\eta_1+\chi_2\eta_2-in R t_1'-in R t_2'}
\label{repr1}\\
&\times\int d\bar X'd X'd\bar \Psi d \Psi
\exp\{(Q\tilde \Psi,\tilde \Psi^*)\},
\notag
\end{align}
where 
\begin{align}\label{F}
P(X,\Psi, R)=\int \exp\Big\{&\chi_1(\bar X_1',\Psi_1)(R/n)^{1/2}+\eta_1(X_2',\bar \Psi_2)(R/n)^{1/2}+\chi_2(\bar X_2',\Psi_2)(R/n)^{1/2}\\ \notag
&+\eta_2(X_1',\bar \Psi_1)(R/n)^{1/2}+\chi_1\eta_1+\chi_2\eta_2\Big\}\, d\eta d\chi 
\end{align}
with $d\eta d\chi  =d\eta_1 d\chi_1d\eta_2 d\chi_2 $. 
We also denoted $\tilde\Psi,\tilde \Psi^*$  super-vectors of the form
\begin{align*}
\tilde \Psi=\left(\begin{matrix}\Psi_1\\ \Psi_2\\ X_1'\\ X_2'\end{matrix}\right),\quad \tilde \Psi^*=\left(\begin{matrix}\bar \Psi_1\\ \bar \Psi_2 \\ \bar X_1' \\ \bar X_2'\end{matrix}\right)
\end{align*}
with Grassmann vectors $\Psi_{1,2},\bar \Psi_{1,2}$ and complex vectors $X'_{1,2},\bar X'_{1,2}$. $Q$ is $4n\times 4n$ super-matrix of the form
\begin{align*}
Q=&\left(\begin{array}{cc}A(u'-\varepsilon_1)&\hat \chi(R/n)^{1/2}\\ \hat \eta (R/n)^{1/2}&R\cdot B(t_{1\varepsilon},t_{2\varepsilon})\end{array}\right),
\quad t_{1\varepsilon}=t_1'+i\varepsilon\frac{r_1}{r_2},\quad
t_{2\varepsilon}=t_2'+i\varepsilon\frac{r_2}{r_1}
\end{align*}
and $2n\times 2n$ block matrices  $A, B$ have complex coefficients, while $\hat \chi,\hat \eta$ are diagonal matrix with Grassmann coefficients
$ \chi_1, \chi_2,\eta_1,\eta_2$:
\begin{align*}
 A(u)=&\left(\begin{array}{cc}u I_{n}&i(A_0-z)\\ i(A_0-z)^*&\bar u I_{n}\end{array}\right),
\quad B(t_1,t_2)=\left(\begin{array}{cc}it_1I_{n}&i(A_0-z)\\ i(A_0-z)^*&it_2I_{n}\end{array}\right),\\
\hat \chi=&\left(\begin{array}{cc}\chi_1I_{n}&0\\ 0&\chi_2I_{n}\end{array}\right),\quad 
\hat \eta=\left(\begin{array}{cc}\eta_2I_{n}&0\\ 0&\eta_1I_{n}\end{array}\right).
\end{align*}
It is easy to see that
\begin{align}\label{detA}
&\det A=\det ((A_0-z)(A_0-z)^*+|u|^2),\\
\notag &\det B=\det ((A_0-z)(A_0-z)^*-t_1t_2).
\end{align}
Notice that to simplify formulas here and below we drop the variables dependence in the notations $A(u)$, $B(t_1,t_2)$.

Now we can integrate with respect to
$d\bar X'd X'd\bar \Psi d \Psi$ using (\ref{G_comb}):
\begin{align}\label{sdetQ}
\int dX'd\bar X'd\Psi d\bar \Psi&
\exp\{(Q\tilde \Psi,\tilde \Psi^*)\} =\Sdet^{-1}Q=\frac{\det A}{\det RB}\, \Sdet^{-1}(1+\tilde Q)\\ \notag
=&\frac{\det A}{\det (RB)}\,\exp\{-\Str\log(1+\tilde Q)\}
=\frac{\det A}{R^{2n}\det B}\,\exp\Big\{\frac{1}{2}\,\Str\,\tilde Q^2+\frac{1}{4}\,\Str\,\tilde Q^4\Big\},
\end{align}
where
\begin{align*}
\tilde Q:=\left(\begin{array}{cc}0&A^{-1}\hat \chi(R/n)^{1/2}\\ -R^{-1}B^{-1}\hat \eta (R/n)^{1/2}&0\end{array}\right),\quad
\tilde Q^2=\dfrac{1}{n}\left(\begin{array}{cc}-A^{-1}\hat \chi B^{-1}\hat \eta&0\\ 0&-B^{-1}\hat \eta A^{-1}\hat \chi\end{array}\right).
\end{align*}
Here we used that 
\[\Str\,\tilde Q=\Str\,\tilde Q^3=0,\quad\tilde Q^{p}=0, \quad p\ge 5.
\] 
Define
\begin{align*}
I=&\int d\eta d\chi \,e^{\chi_1\eta_1+\chi_2\eta_2}\exp\{\frac{1}{2}\Str\,\tilde Q^2+\frac{1}{4}\Str\,\tilde Q^4\}\\
=&\int d\eta d\chi \, e^{\chi_1\eta_1+\chi_2\eta_2}\exp\{\frac{1}{n}\Tr A^{-1}\hat \chi B^{-1}\hat \eta-\frac{1}{2n^2}\Tr (A^{-1}\hat \chi B^{-1}\hat \eta)^2\}\\
=&\int d\eta d\chi \, e^{\chi_1\eta_1+\chi_2\eta_2}(1+\frac{1}{n}\Tr A^{-1}\hat \chi B^{-1}\hat \eta
+\frac{1}{2n^2}(\Tr A^{-1}\hat \chi B^{-1}\hat \eta)^2-\frac{1}{2n^2}\Tr (A^{-1}\hat \chi B^{-1}\hat \eta)^2).
\end{align*}
Observe that
\begin{align*}
A^{-1}\hat \chi B^{-1}\hat \eta=\left(\begin{array}{cc}A^{-1}_{11}B^{-1}_{11}\chi_1\eta_2+A^{-1}_{12}B^{-1}_{21}\chi_2\eta_2
&A^{-1}_{11}B^{-1}_{12}\chi_1\eta_2+A^{-1}_{12}B^{-1}_{22}\chi_2\eta_2\\ 
A^{-1}_{21}B^{-1}_{11}\chi_1\eta_1+A^{-1}_{22}B^{-1}_{21}\chi_2\eta_1
&A^{-1}_{21}B^{-1}_{12}\chi_1\eta_1+A^{-1}_{22}B^{-1}_{22}\chi_2\eta_1\end{array}\right).
\end{align*}
Hence
\begin{align}\label{I}
I=&\int d\eta d\chi \,e^{\chi_1\eta_1+\chi_2\eta_2}(1+\frac{1}{n}\chi_2\eta_2\Tr A^{-1}_{12}B^{-1}_{21})(1+
\frac{1}{n}\chi_1\eta_1\Tr A^{-1}_{21}B^{-1}_{12})\\ \notag
&-\frac{1}{n^2}\chi_1\eta_1\chi_2\eta_2(\Tr A^{-1}_{11}B^{-1}_{11}\Tr A^{-1}_{22}B^{-1}_{22}+\Tr A^{-1}_{12}B^{-1}_{22}A^{-1}_{21}B^{-1}_{11}-
\Tr A^{-1}_{11}B^{-1}_{22}A^{-1}_{22}B^{-1}_{21})\\ \notag
=&(1+\frac{1}{n}\Tr A^{-1}_{12}B^{-1}_{21})(1+
\frac{1}{n}\Tr A^{-1}_{21}B^{-1}_{12})\\ \notag
&-\frac{1}{n^2}\Tr A^{-1}_{11}B^{-1}_{11}\,\Tr A^{-1}_{22}B^{-1}_{22}-\frac{1}{n^2}(\Tr A^{-1}_{12}B^{-1}_{22}A^{-1}_{21}B^{-1}_{11}-
\Tr A^{-1}_{11}B^{-1}_{12}A^{-1}_{22}B^{-1}_{21}).
\end{align}
Using the standard Schur inversion formula for  block matrices, we get
\begin{align}\label{A^-1}
A^{-1}=\left(\begin{array}{cc}\bar u G(u\bar u)&-iG(u\bar u)(A_0-z)\\ -i(A_0-z)^*G(u\bar u)&u\tilde G(u\bar u)\end{array}\right),\\
B^{-1}=\left(\begin{array}{cc}it_2 G(-t_1t_2)&-iG(-t_1t_2)(A_0-z)\\ -i(A_0-z)^*G(-t_1t_2)&it_1\tilde G(-t_1t_2)\end{array}\right),
\notag\end{align}
where $G$ is defined in (\ref{L,G}), and 
\begin{align}\label{ti-G}
\tilde G(x):=((A_0-z)^*(A_0-z)+x)^{-1}.
\end{align}
By (\ref{A^-1}) we have
\begin{align*}
\Tr A^{-1}_{21}B^{-1}_{12}=&\Tr A^{-1}_{12}B^{-1}_{21}=-\Tr G(u\bar u)(A_0-z)(A_0-z)^*G(-t_1t_2)\\
=&-\Tr G(-t_1t_2)+|u|^2\Tr G(u\bar u)G(-t_1t_2);\\
\Tr A^{-1}_{11}B^{-1}_{11}\Tr A^{-1}_{22}B^{-1}_{22}=&-t_1t_2|u|^2(\Tr G(u\bar u)G(-t_1t_2))^2;\\
\Tr A^{-1}_{12}B^{-1}_{22}A^{-1}_{21}B^{-1}_{11}=&t_1t_2\Tr G(u\bar u)(A_0-z)\tilde G(-t_1t_2)(A_0-z)^*G(u\bar u)G(-t_1t_2)\\
=&t_1t_2\Tr G^2(-t_1t_2)(A_0-z)(A_0-z)^*G^2(u\bar u)\\
=&t_1t_2\Tr G^2(-t_1t_2)G(u\bar u)-t_1t_2|u|^2\Tr G^2(-t_1t_2)G^2(u\bar u);\\
\Tr A^{-1}_{11}B^{-1}_{12}A^{-1}_{22}B^{-1}_{21}=&-|u|^2\Tr G(u\bar u)G(-t_1t_2) (A_0-z)\tilde G(u\bar u)(A_0-z)^*G(-t_1t_2)\\
=&-|u|^2\Tr G(u\bar u)G^2(-t_1t_2) +|u|^4\Tr G^2(u\bar u)G^2(-t_1t_2).
\end{align*}
Here we  used the relation
\[\tilde G(x)(A_0-z)^*=(A_0-z)^*G(x).\]
Substituting this to (\ref{I}), we obtain
\begin{align}\notag
I=&\Big(1-\frac{1}{n}\Tr G(-t_1t_2)+\frac{|u|^2}{n}\Tr G(u\bar u)G(-t_1t_2)\Big)^2+t_1t_2|u|^2\Big(\frac{1}{n}\Tr G(u\bar u)G(-t_1t_2)\Big)^2\\
&-\frac{|u|^2+t_1t_2}{n^2}\Big(\Tr G(u\bar u)G^2(-t_1t_2)-|u|^2\Tr G^2(u\bar u)G^2(-t_1t_2)\Big).\label{I_1}
\end{align}
Finally, let us change the variables in the integral with respect to $u_1',u_2',t_1',t_2'$ of (\ref{repr1}) as
\[u=u'-\varepsilon_1,\quad t_{1}=t_1'+i\varepsilon\frac{r_1}{r_2},\quad
t_{2}=t_2'+i\varepsilon\frac{r_2}{r_1}.\]
Notice, that in the case of $t_1,t_2$ this means that we move integration from the lines $t'_{1,2}\in\mathbb{R}$ to the lines
$t_1\in L_1=\mathbb{R}+i\varepsilon_1\frac{r_1}{r_2}$, $t_2\in L_2=\mathbb{R}+i\varepsilon_1\frac{r_2}{r_1}$. This can be done since the function
$\mathcal{L}_n(-t_1t_2)$ is analytic in $t_1,t_2$, if $\Im t_1> 0$, $\Im t_2> 0$.  Moving  the integration with respect to $t_1,t_2$ 
 to $L$, gathering together (\ref{repr1}), (\ref{detA}), (\ref{sdetQ}), (\ref{I_1}), and differentiating with respect
to $\varepsilon$, we obtain (\ref{p1.1}).

$\square$

\section{Proof of Theorem \ref{t:1}.}\label{s:t1}
Now we perform an asymptotic analysis of (\ref{p1.1}) in the regime $\hbox{dist}\{z,\partial D\}\sim n^{-1/2}$, $\varepsilon\sim n^{-3/4}$
(recall that we consider only $\omega\in \Omega_{\epsilon_0/2}^{(1)}\cap \Omega_{\epsilon_0/2}^{(2)}\cap \Omega^{(3)}$ (see conditions (A3) -- (A4)).

Let us change the variables in (\ref{p1.1}):
\begin{align}\label{ch1}
(t_1,t_2)\to(s,t),\quad s=\frac{t_1-t_2}{2},\quad  t=\frac{t_1+t_2}{2},\quad s\in\mathbb{R},\quad t\in L,
\end{align}
and 
\begin{align}\label{ch2}
(r_1,r_2)\to (v, R),\quad v=r_1-r_2,\quad R=r_1r_2
\end{align}
Jacobian of the first change is $J_1=\frac{1}{2}$, and for the second change it is
\[J_2=\frac{1}{r_1+r_2}=\frac{1}{(v^2+4R)^{1/2}},\]
so (\ref{p1.1}) in new variables takes form
 \begin{align}\label{p1.1_new}
T(z,\varepsilon)=&\frac{n^3}{4\pi^3\varepsilon}\int_{-\infty}^\infty (u_1+\varepsilon)du_1 d u_2  \int_{-\infty}^\infty \dfrac{ds\, dv}{(v^2+4R)^{1/2}} \int_{L} dt \int_{0}^\infty R \,dR 
\\&\times \varphi(u_1,  u_2,  t+s, t-s)
\exp\{nF_1(u_1,u_2)\}\exp\{n\tilde F_2(s,t,v,R)\}\notag
\end{align}
where
\begin{align}\label{tilde_F}
\tilde F_2(s,t,v,R)=-\mathcal{L}_n(s^2-t^2)-(R+it+\varepsilon)^2-(t-i\varepsilon)^2-\varepsilon v^2.
\end{align}
Let us move the integration with respect to $t$ to the contour 
\begin{align}\label{L}
&L= L_1\cup L_1'\cup L_2\cup L_2',\\ \notag
& L_{1,2}=\{t=iu_*\pm e^{\pm i\pi/4}\tau, \tau\in [0,C_0]\},  \\
& L_{1,2}'=\{t=iu_*\pm e^{\pm i\pi/4}C_0\pm\tau,\,\tau>0\},\notag
\end{align}
where $C_0$ is sufficiently big to provide the inequality
\begin{align}\label{C_0}
\log C_0^2>\mathcal{L}(0)+2,
\end{align} 
and 
\begin{align}\label{eq_u}
u_*=n^{-1/4}.
\end{align}

To show the possibility of such contour shift, we notice that  for any fixed $s$ the function $\mathcal{L}_n(s^2-t^2)$ is analytic in $t$ in 
the domain $\{t: \Im t>0\}$; moreover, for any fixed $s$
\[-\Re\mathcal{L}_n(s^2-t^2)<-\frac{1}{2}\log(|t|^2+1), \quad |t|\to \infty.\] 
Continuing the contours deformation, 
we deform the $R$-contour for any fixed $t\in L$  as  follows (this deformation is possible in view of (\ref{tilde_F})): 
\begin{align}\label{R(t)}
&\mathcal{R}(t)=\mathcal{R}_1(t)\cup \mathcal{R}_2(t), \\ \notag
&\mathcal{R}_1(t)=\{R: R=e^{i\theta(t)}\rho,\,0\le \rho\le |it+\varepsilon|\},\quad \theta(t)=\mathrm{arg}(-it-\varepsilon)\\
&\mathcal{R}_2(t)=\{R: R=-it-\varepsilon+\rho,\,\,\rho>0\},\notag
\end{align}
Below we will need the following straightforward inequalities: 
\begin{align}\label{in_R}
-\Re(R+it+\varepsilon)^2
\le\left\{\begin{array}{lll}-\cos (2\theta(t))\cdot(|it+\varepsilon|-\rho)^2\le 0,&t\in L_1\cup L_2, &R\in \mathcal{R}_1(t),\\
-\rho^2<0,&t\in L_1\cup L_2,& R\in \mathcal{R}_2(t),\\
%
\Re(t-i\varepsilon)^2+O(u_*),&t\in L_1'\cup L_2', &R\in \mathcal{R}_1(t),\\
-\rho^2<0,&t\in L_1'\cup L_2',& R\in \mathcal{R}_2(t),\end{array}\right.
\end{align}
Notice that the term $O(u_*)$ appears for only for $0\le\tau\le u_*$ in (\ref{L}). 
For $ t\in L_1'\cup L_2'$ we have
\begin{align}\label{in_L'}
-\Re\mathcal{L}_n(s^2-t^2)=&-\Re\mathcal{L}_n(s^2-(\pm C_0e^{\pm i\pi/4}\pm\tau)^2) +O(u_*)\\
=&-\frac{1}{2}\int \log\Big( (\lambda+s^2-\sqrt 2C_0\tau-\tau^2)^2+(C_0^2+\sqrt2 C_0\tau)^2\Big)d\nu_{n,z}(\lambda)+O(u_*)\notag\\
\le&-\log C_0^2+O(u_*)\le -(\mathcal{L}_n(0)+1).\notag
\end{align}
Here and below we denote by $\nu_{n,z}(\lambda)$ an empirical spectral measure of $(A_0-z)(A_0-z)^*$.

We are going to integrate first  with respect to $s$, then with respect to $R$, and then with respect to $u_1,u_2,t,v$.

To integrate with respect to $s$, observe that for $t\in L_1\cup L_2$
\[\frac{d}{ds}  \tilde F_2(s,t,v,R)\Big|_{s=0}=0,\]
 and
\[ 
\frac{d}{d(s^2)} \Re \tilde F_2(s,t,v,R)=-\frac{1}{2}\frac{d}{d(s^2)} \int\log\Big((\lambda+s^2+\sqrt 2u_*\tau+u_*^2)^2+(\tau^2+\sqrt 2u_*\tau)^2\Big)
d\nu_{n,z}(\lambda)<0.
\]
For $t\in L_1'\cup L_2'$ under condition  (\ref{C_0}) for $C_0$ we have by  (\ref{in_R}) and (\ref{in_L'})
\[ n\Re \tilde F_2(s,t,v,R)<-(n-2)(\mathcal{L}_n(0)+1)-\int \log\Big( (\lambda+s^2-\sqrt 2 C_0\tau-\tau^2)^2+C_0^4\Big)d\nu_{n,z}(\lambda).
\]
Then a saddle-point method with respect to $s$  gives 
\begin{align}\label{saddle_s}
\varphi(u_1,u_2,t+s,t-s)e^{n\tilde F_2(s,t,v,R)}&\to \sqrt{\pi/(n F_{s})}\varphi(u_1,u_2,t,t)e^{n\tilde F_2(0,t,v,R)}(1+O(n^{-1})),\\
F_{s}:=n^{-1}\Tr G(-t^2).
\notag\end{align}
Notice $F_s\to 1$ as $t\to 0$ (see (\ref{L,G}), (\ref{dD}) and recall $\hbox{dist}\{z,\partial D\}=\tilde\delta^2 n^{-1/2}$)

To integrate with respect to $R$, 
we would like to restrict the integration domain to
\begin{align}\label{restr_t}
|t|\le n^{-1/4}\log n.
\end{align}
To this end, observe that for any $t\in L_1\cup L_2$
\begin{align}\label{t^2}
(it+\varepsilon)^2=&-(t-i\varepsilon)^2=-(\pm\tau e^{\pm i\pi/4}+i(u_*-\varepsilon))^2\\
=&\mp  i(\tau^2+\sqrt 2(u_*-\varepsilon))+(u_*-\varepsilon)^2+\sqrt 2(u_*-\varepsilon)\tau.
\notag\end{align}   
Hence, using also (\ref{in_R}), for $t\in L_1\cup L_2$ we get
\begin{align}\label{ReF}
 \Re \tilde F_2(0,t,v,R)<& \Re \Big(-\mathcal{L}_n(-t^2)-(t-i\varepsilon)^2\Big)-\varepsilon v^2\\
 = &-\frac{1}{2}\int\log\Big((\lambda+\sqrt 2u_*\tau+u_*^2)^2+(\tau^2+\sqrt 2u_*\tau)^2\Big)d\nu_{n,z}(\lambda)\notag
\\& +((u_*-\varepsilon)^2+(u_*-\varepsilon)\sqrt 2\tau)-\varepsilon v^2.
\notag \end{align}
Set
\begin{align}\label{F_2}
\tilde F(t):=-\mathcal{L}_n(-t^2)-(t-i\varepsilon)^2.
\end{align}
 Using (\ref{ReF}) it is  straightforward to show that 
 \begin{align}\label{F_2'}
\Re \tilde F'(t)=0\Rightarrow |t|^3< C u_*\Leftrightarrow |t|< C_1n^{-1/12}.
\end{align}
Let us expand  $\tilde F(t)$ for  $|t|< C_1n^{-1/12}$ as follows: 
 \begin{align*}
  \tilde F(t)=-\mathcal{L}(0)+\frac{t^2}{n}\Tr G(0)+\frac{t^4}{2n}\Tr G^2(0)-(t-i\varepsilon)^2+O(t^6).
 \end{align*}
Denote by $z_*$ the nearest to $z$  point of the boundary $\partial D$. Since we assume that $\partial D$ is smooth near $z$ 
we have
\[z-z_*=|z-z_*|\,|\triangledown\mathcal{F}(z_*)|^{-1}\triangledown\mathcal{F}(z_*)+O(\delta^4), \quad |\triangledown\mathcal{F}(z_*)|\not=0.
\]
with $\mathcal{F}$ defined in (\ref{cal-F}).
Then, taking into account that we consider $\omega\in \Omega_{\epsilon_0/2}^{(2)}$ (see assumption (A3)), we obtain
\begin{align}\notag
n^{-1}\Tr G(0)=&n^{-1}\Tr Y_0^{-1}(z_*)+n^{-1}\Big(\Tr Y_0^{-1}(z)-\Tr Y_0^{-1}(z_*)\Big)\notag\\
=&1+O(n^{-1/2-d_0})+n^{-1/2}\tilde\delta^2 k+O(n^{-1}), \notag\\
 k=&|\triangledown\mathcal{F}(z_*)|=\Big|n^{-1}\Tr G^2(0)(A_0-z)\Big|(1+o(1)),\label{ti-de_1}
\end{align}
where we used also the assumption of Theorem \ref{t:1} that $|z-z_*|=n^{-1/2}\tilde \delta^2$.

Hence
\begin{align}\label{ReF_2}
\tilde F(t)=&-\mathcal{L}(0)+t^2(1+n^{-1/2}k\tilde\delta^2)-(t-i\varepsilon)^2+\frac{c_2t^4}{2}+O(t^2n^{-1}) +O(t^6) \\
=&-\mathcal{L}(0)+\frac{c_2t^4}{2}+t^2k\tilde\delta^2n^{-1/2}(1+o(1))+2it\varepsilon+\varepsilon^2+ O(t^6) +O(n^{-1}t^2),
\notag\end{align}
where 
\begin{align}\label{c_2}
c_2=n^{-1}\Tr G^2(0).
\end{align}
 Thus for $t\in L_1\cup L_2$ and $|t|\le C_1n^{-1/12}$
 \begin{align*}
 \Re \tilde F(t)+\mathcal{L}(0)\le
 &\frac{c_2}{2}(u_*^4+2\sqrt 2u_*^3\tau-2\sqrt 2u_*\tau^3-\tau^4)-k\tilde\delta^2n^{-1/2}(u_*^2+\sqrt 2u_*\tau) \\&
 +O(n^{-1/6})\tau^4 +O(n^{-3/4})\tau.
 \end{align*}
Replacing $\tau\to n^{-1/4}\tilde\tau$, we get
 \begin{align*}
 \Re \tilde F(\tau)+\mathcal{L}(0)\le n^{-1}
 &\Big(-\frac{c_2}{3}\tilde \tau^4 +q_3\tilde \tau^3+q_2\tilde \tau^3+q_1\tilde \tau+q_0\Big)
 \end{align*}
 with some bounded $q_{0,1,2,3}$.  By (\ref{F_2'})  $\tilde F(t)$ decays  for
 $|\tau|>C_1n^{-1/12}$ and for \\ $n^{-1/4}\log n\le|\tau|\le C_1n^{-1/12}$  the above relation implies
 \begin{align*}
 &\Re \tilde F(t)\le -\mathcal{L}(0) -\frac{c_2}{4}\tau^4<-\mathcal{L}(0)-Cn^{-1}\log^4n,\quad t\in L_1\cup L_2.
 \end{align*}
 For  $t\in L_1'\cup L_2'$ with $C_0$  satisfying (\ref{C_0}) we have by (\ref{in_R}) and (\ref{in_L'})
\begin{align*}
 \Re \tilde F_2(0,t,v,R)&<-(\mathcal{L}_n(0)+1).
\end{align*} 
Thus finally we obtain the inequality (see (\ref{in_L'}))
\begin{align*}
n\Re \tilde F_2(0,\tau,v,R)-n\tilde F_2(0,0,0,0)\le -C\log^4n-\log(C_0^2+\tau C_0)^2-R^2-2itR-\varepsilon(2R+v^2).
\end{align*}
Thus  we conclude that we can restrict the integration with respect to $t$ to the domain (\ref{restr_t}).

Similarly, if we change  $u_1=|u|\cos\phi$, $u_2=|u|\sin\phi$,  $du_1du_2=|u|d|u|d\phi$, then
we can restrict the integration with respect to $|u|$ by $|u|\le n^{-1/4}\log n$.

Since  the $R$-dependent part of $\tilde F_2$ of (\ref{tilde_F}) for any $|t|<n^{-1/4}\log n$ has the form
\[
\tilde F_R=-(R+it+\varepsilon)^2,
\]
we can integrate over $R$. Notice that for $|t|<n^{-1/4}\log n$ and $R\in \mathcal{R}_1(t)$ 
\begin{align*}
\tilde F_R=-\cos (2\theta(t))\cdot (\rho-|it+\varepsilon|)^2
\end{align*}
(see (\ref{R(t)}) for the parametrization of $\mathcal{R}_1(t)$). In addition, since $-it-\varepsilon=u_*-\varepsilon\mp i\tau e^{\pm i\pi/4}$, we have
\begin{align*}
\cos (2\theta(t))=&1-2\sin^2\theta(t)=1-\frac{\tau^2}{\tau^2+(u_*-\varepsilon)^2+\sqrt 2 (u_*-\varepsilon)\tau}\\
\ge &\frac{(u_*-\varepsilon)}{(\tau^2+(u_*-\varepsilon)^2+\sqrt 2 (u_*-\varepsilon)\tau)^{1/2}}=\frac{(u_*-\varepsilon)}{|it+\varepsilon|}>\frac{C}{\log n}.
\end{align*}
Here we have (\ref{eq_u}) and (\ref{restr_t}). Moreover, for $R\in \mathcal{R}_2(t)$
\[
\tilde F_R=-\rho^2.
\]
Hence, the main contribution to the $R$-integral is given by the  the saddle-point $R=-it-\varepsilon$, and 
we obtain after integration with respect to $R$:
\begin{align}\label{R=0}
\frac{Re^{n\tilde F(t)-n(R+it+\varepsilon)^2-n\varepsilon v^2+n\varepsilon^2} }{(v^2+4R)^{1/2}}\to 
\sqrt{\pi/n}\frac{-(it+\varepsilon)}{(v^2-4(it+\varepsilon))^{1/2}}e^{n\tilde F(t)-n\varepsilon v^2}(1+O(n^{-1/4}))
\end{align}
with $\tilde F(t)$ of (\ref{F_2}). 

Relations (\ref{ti-de_1}), (\ref{ReF_2}),   and  (\ref{restr_t}) yield
\begin{align*}
\tilde F(t)-\varepsilon v^2=
-\mathcal{L}(0)+\frac{c_2t^4}{2}+t^2n^{-1/2}k\tilde\delta^2+2it\varepsilon -\varepsilon v^2+O(n^{-3/2}\log^6n) 
\end{align*}
Moreover, for small $x$
\begin{align}\label{exp_G}
n^{-1}\Tr G(x)=&n^{-1}\Tr G(0)-x n^{-1}\Tr G^2(0)+O(x^2),\\
n^{-1}\Tr G^2(x)=&n^{-1}\Tr G^2(0)+O(x)=:c_2+O(x).
\notag\end{align}
which combined with (\ref{ti-de_1})   and  (\ref{restr_t})  implies for $\varphi$ of (\ref{p1.2})
\begin{align}\label{exp_phi}
\varphi(|u|\cos\phi,|u|\sin\phi,t,t)=&(-n^{-1/2}k\tilde\delta^2+(-t^2+|u|^2)c_2)^2+t^2|u|^2c_2^2\\
&+O(|t|^6+|u|^6)+O(n^{-1}(|t|^2+|u|^2))\notag
\end{align}
with $c_2$ of (\ref{c_2}).

Now  let us make the change of variables
\begin{align*}
|u|=&n^{-1/4}\tilde u,\quad\\
 t=&n^{-1/4}\tilde t, \quad \tilde t\in \tilde L_1\cup\tilde L_2,\quad
\tilde L_{1,2}=\{z:z=\pm\tilde \tau e^{\pm i\pi/4},\,\,\tilde \tau>0\},\\
v^2= &(-in^{-1/4}\tilde t-\varepsilon)\tilde v^2,
\end{align*}
where the sign $\pm$ corresponds to $\tilde t\in \tilde L_1$ and $\tilde t\in \tilde L_2$.
Then  we have
\begin{align*}
\varphi(u_1,u_2,t_1,t_2)\to&n^{-1}(-k\tilde\delta^2+(\tilde u^2-\tilde t^2)c_2)^2+\tilde t^2\tilde u^2c_2^2+O(n^{-3/2}),\\
nF_1(u_1,u_2)\to &n\mathcal{L}(0)+k\tilde\delta^2\tilde u^2-\frac{c_2}{2}\tilde u^4+2\tilde\varepsilon\tilde u \cos\theta+O(n^{-1/2}),\\
n\tilde F(t)-\varepsilon v^2\to&-n\mathcal{L}(0)+k\tilde\delta^2\tilde t^2+\frac{c_2}{2}\tilde t^4+2i\tilde t\tilde\varepsilon-\tilde\varepsilon \tilde v^2.
\end{align*}
Finally we obtain
\begin{align*}
T=\frac{n^{1/2}}{4\pi^2\tilde\varepsilon}\sum_{\pm}&\Big(\mp\int _0^\infty \tilde u^2e^{-c_2\tilde u^4/2+k\tilde\delta^2\tilde u^2}d\tilde u
\int _{0}^\infty\tilde \tau e^{-c_2\tilde \tau^4/2\pm ik\tilde \delta^2\tilde \tau^2+2i\tilde\varepsilon\tau e^{\pm i\pi/4}}d\tilde \tau\\
&\times\int_{0}^{2\pi}\cos\theta e^{-\varepsilon\tilde u \cos\theta }d\theta 
\int_{-\infty}^\infty\frac{e^{-e^{\mp i\pi/4}\tilde \varepsilon\tilde\tau \tilde v^2}dv}{(\tilde v^2+4)^{1/2}}\\
&\times\left((-k\tilde\delta^2+(\tilde u^2\mp i\tilde \tau^2)c_2)^2 \pm i\tilde \tau^2\tilde u^2c_2^2+O(n^{-3/2})\right) \Big).
\end{align*}
$\square$


\section{Proof of Theorem \ref{t:2}}\label{s:t2}
Here we perform an asymptotic analysis of $T(z,\varepsilon)$ of (\ref{p1.1}) in the regime $\hbox{dist}\{z,\partial D\}\sim 1$, $\varepsilon\sim n^{-1}$ corresponding
to (\ref{SST}) bound. Again, we consider only $\omega\in \Omega_{\epsilon_0/2}^{(1)}\cap \Omega_{\epsilon_0/2}^{(2)}\cap \Omega^{(3)}$ (see conditions (A3) -- (A4)).

We again  change the variables as in (\ref{ch1}) and (\ref{ch2}) and obtain (\ref{p1.1_new}).

Denote by $u_*$  a positive solution of the equation
\begin{align}\label{eq_u*}
n^{-1}\Tr G(u^2_*)=1
\end{align}
with $G(x)$ defined in (\ref{L,G}).

Set 
\begin{align}\label{ti_F21}
\tilde F(t)=-\mathcal{L}_n(-t^2)-t^2.
\end{align}
\begin{lemma}\label{l:count}
One can chose a sufficiently small $0<\kappa<u_*$  and sufficiently big $C_0$  satisfying
(\ref{C_0}), such that  there exists a contour $L_1\subset \{z:\Re z>0\, \wedge\,\mathrm{arg}\,z\ge\pi/4\}$ and a constant $\sigma>0$ depending on
$L_1$ with the following conditions:
\begin{align}\label{count.1}
&iu_*+\kappa\in L_1,\quad iC_0+ C_0\in L_1,\\
&\Re \tilde F(t)\le \Re \tilde F(iu_*)-\sigma,\quad t\in L_1.
\label{count.2}\end{align}
\end{lemma}
\textit{Proof.}
Since  it is easy to check that $x=0$ is a minimum point of $\Re\tilde F(iu_*+x)$ ($x\in \mathbb{R}$), we conclude that $\Re\tilde F''(iu_*)<0$
and  we can choose $\kappa$ sufficiently small to provide that
 $\Re\tilde F$ decreases  on the segment $[iu_*,iu_*+\kappa]$ and hence the segment is situated between two level lines  of $\Re\tilde F$: 
 $\ell_1$ and $\ell_2$ which intersect in $iu_*$ ( $\ell_1$ is an upper level line). 
 Here and below we denote $[a,b]$ the segment with edge points $a,b\in \mathbb{C}$. Notice that
  since  $t=u_*$ is a maximum point of $\Re\tilde F(it)$, $\ell_1$ and $\ell_2$ cannot intersect the imaginary axis at $it\not= iu_*$.
  Let us also mention that $\ell_1$ and $\ell_2$ cannot form loops or intersect with each other in the upper half-plane since $\Re \tilde F$
  is a harmonic function.
  
Recall that $\nu_{n,z}$ is the normalized counting measure of $(A_0-z)(A_0-z)^*$. Taking into account that
\begin{align}
\Re \tilde F(\tau e^{i\pi/4})=-\frac{1}{2}\int\log(\lambda^2+\tau^4)d\nu_{n,z}(\lambda)
\end{align}
decreases, as $\tau$ grows, it suffices to prove that $\ell_2$ intersects the ray $\ell_3=\{z=te^{i\pi/4}\}$ at some point $t_*e^{i\pi/4}$.
Indeed, in this case, 
since $\ell_1$ cannot intersect $\ell_3$ and $l_2$ cannot intersect $\ell_3$ twice, we can choose any curve, satisfying (\ref{count.1}), lying between $\ell_1$ and $\ell_2$, and above
or on $\ell_3$, and such that the distances between $L_1$ and
$\ell_1$ and $L_1$ and $\ell_2$ are not zero.

Hence it suffices to prove that $\ell_2$ intersects $\ell_3$. Choose
$ \lambda_*$ such that
\[\lambda_*^2\ge\sup \{\lambda\in\mathrm{supp}\,\nu_{n,z}\}.
\]
Consider $\ell_4=\{i\lambda_*+\tau,\,\tau>0\}$.
Then since
\begin{align*}
\frac{d}{d(\tau^2)}
\Re\tilde F(i\lambda_*+\tau)=
-\int\frac{\lambda_*^2+\tau^2-\lambda}{(\lambda_*^2+\lambda-\tau^2)^2+4\lambda_*^2\tau^2} d\nu_{n,z}(\lambda)-1\le-1,
\end{align*}
 we conclude that $\Re\tilde F$ decreases at $\ell_4$. Since $\ell_1$ cannot intersect $\ell_3$, it must intersect $\ell_4$. But then
 $\ell_2$ cannot intersect $\ell_4$, and therefore it must intersect $\ell_3$.
$\square$

Now we take $L_1$ given by Lemma \ref{l:count}, take $L_2$ to be symmetric to $L_1$ with respect to the imaginary axis, and
consider the contour 
\begin{align}\label{contL}
L=&L_0\cup L_1\cup L_2\cup L_1'\cup L_2',\\
L_0=&[(iu_*-\kappa),(iu_*+\kappa)],\quad
 L'_{1,2}=\{(iC_0\pm C_0)\pm\tau,\,\tau>0\},\notag
\notag\end{align}
As in the case of Theorem \ref{t:1}, we move the integration  with respect to $t$ from the real line to $L$.


Then for any fixed $t\in L$ we move  the contour of integration  with respect to $R$ to $\mathcal{R}(t)$ of (\ref{R(t)}) with $\varepsilon=0$.
Notice that inequalities (\ref{in_R}) with $\varepsilon=0$ are still valid for $L_1\cup L_2$ and for $t\in L_1'\cup L_2'$ (without $O(u_*)$ term),
and the inequality (\ref{in_L'}) should be replaced with
\begin{align}\label{in_L'1}
-\Re\mathcal{L}_n(s^2-t^2)<-(\mathcal{L}_n(u^2_*)+1),\quad t\in L_1'\cup L_2'.
\end{align}
To restrict the integration with respect to $s$, notice that for $t\in L_0$
\begin{align*}
-&\Re\mathcal{L}_n(s^2-(iu_*+\tau)^2)=-\frac{1}{2}\int \log ((\lambda+s^2+u^2_*-\tau^2)^2+4u^2_*\tau^2) d\nu_{n,z}(\lambda),
\end{align*}
and thus, since $|\tau|\le\kappa<u_*$, for $\tilde F_2$ from (\ref{tilde_F}) we get
\begin{align*}
 &\frac{d}{d(s^2)}
\Re\tilde F_2=
-\int\frac{\lambda+s^2+u^2_*-\tau^2}{(\lambda+s^2+u^2_*-\tau^2)^2+4u^2_*\tau^2} d\nu_{n,z}(\lambda)< 0,
\end{align*} 
Moreover, for $t\in L_1\cup L_2$ we have
\begin{align}\label{expr_L'}
-\Re\mathcal{L}_n(s^2-\Re(t^2)-i\Im (t^2))=-\frac{1}{2}\int \log  ((\lambda+s^2-\Re(t^2))^2+\Im^2(t^2)) d\nu_{n,z}(\lambda),
\end{align}
and since $\Re(t^2)\le 0$ (recall that  by Lemma \ref{l:count} $L_1$ is above $\ell_3$), one can see that
$\Re\tilde F$ is decreasing function of $s^2$, when $t\in  L_1\cup L_2$.

 Combining the above argument with (\ref{in_L'1}) we conclude that there is $C>0$ such that
\begin{align*}
-n\Re\mathcal{L}_n(s^2-t^2)\le- n\mathcal{L}_n(-t^2)-C\log^2 n,\quad | s|>n^{-1/2}\log n,
\end{align*}
and thus we can restrict the integration with respect to $s$ by
\begin{align}\label{restr_s}
|s|\le n^{-1/2}\log n.
\end{align}
Now let us show that we can also restrict the integration with respect to $t$ by
\begin{align}\label{restr_t1}
|t-iu_*|\le (u_*^2n)^{-1/2}\log n.
\end{align}
We would like to notice  that  if $\delta\sim 1$ and $u_*\sim 1$, then  $(nu_*^2)^{-1/2}$ differs from $n^{-1/2}$ only with a constant.
But in the setting of Theorem \ref{t:3}, when $u_*\sim \delta\ll 1$ the factor  $(nu_*^2)^{-1/2}$ gives the correct scaling, so this normalization
is a preparation to the proof of Theorem \ref{t:3}. 

It follows from (\ref{in_R}) and Lemma \ref{l:count}  that for $t\in L_0\cup L_1\cup L_2$, $|t-iu_*|> (u_*^2n)^{-1/2}\log n$ and $s$ satisfying (\ref{restr_s}) we have
\begin{align*}
\Re\tilde F_2(s,t,R,v)\le \Re\tilde F(t)-Cs^2\le \Re \tilde F(iu_*)-Cn^{-1}\log^2n.
\end{align*}
Thus, we prove that the integration with
respect to $t$ can be restricted to (\ref{restr_t1}). Similarly, if we change  $u_1=|u|\cos\phi$, $u_2=|u|\sin\phi$,  $du_1du_2=|u|d|u|d\phi$, then
we can restrict the integration with respect to $|u|$ to 
\begin{align}\label{restr_u}
||u|-u_*|\le (u_*^2n)^{-1/2}\log n.
\end{align}
In addition, using that $\varphi $ of (\ref{p1.2}) does not depend on $R$, we can repeat the argument of Section \ref{s:t1} in order to integrate with respect to $R$ (recall $\varepsilon$ is of order $1/n$):
\begin{align}\label{R=01}
&\int \frac{Re^{-n(R+it+\varepsilon)^2+n\varepsilon^2} }{(v^2+4R)^{1/2}} dR\\ \notag &= 
\sqrt{\pi/n}\frac{-(it+\varepsilon)}{(v^2-4(it+\varepsilon))^{1/2}}(1+O(n^{-1}))=\sqrt{\pi/n}\frac{-it}{(v^2-4it)^{1/2}}(1+O(n^{-1})).
\end{align}
Now, denoting $\Omega_n$ the set of $|u|,s\in\mathbb{R}$ and $t\in iu_*+\mathbb{R}$ satisfying (\ref{restr_s}), (\ref{restr_t}), and (\ref{restr_u}),
  and changing  the variable $v^2=(-4it)\tilde v^2$,  we can write (\ref{p1.1}) in the form
\begin{align}\label{p1.4}
\varepsilon T(z,\varepsilon)=\left\langle |u| \cos\phi\right\rangle (1+O(n^{-1})),
\end{align}
where we introduce the averaging
\begin{align}\label{p1.3}
\left\langle f(\tilde t,\tilde u,\tilde s,\phi,\tilde v)\right\rangle=&\frac{n^{5/2}}{4\pi^{5/2}}\int_{-\pi}^\pi  \,d\phi\int_{-\infty}^\infty \frac{d\tilde v }{(\tilde v^2+1)^{1/2}}\int_{t,s,|u|\in\Omega_n}d|u|  dt ds \,|u|(-it)  f(\tilde t,\tilde u,\tilde s,\phi,\tilde v)
\\&\times\tilde\varphi(|u|, t,s)\cdot e^{n h(|u|)-nh(-it)}e^{-ns^2F_s(t)}e^{-2n\varepsilon|u|\cos\phi}e^{2it(2\tilde v^2+1)\varepsilon n}
\notag 
\end{align}
with
\begin{align*}
&h(x)=\mathcal{L}_n(x^2)-x^2,\quad
F_s(t)=n^{-1}\Tr\, G(-t^2), \\
& \tilde\varphi(|u|, t,s)=\varphi(|u|\cos\phi,|u|\sin\phi, t+s,t-s).
\end{align*}
At the final step we 
make the change of variables in (\ref{p1.1})
\begin{align}\label{ch3}
t=iu_*+(nu_*^2)^{-1/2}\tilde t,\quad s=n^{-1/2}\tilde s,\quad |u|=u_*+(nu_*^2)^{-1/2}\tilde u.
\end{align}
We have
\begin{align}\label{exp_phi_til}
\tilde\varphi(u_*, iu_*,0)=0\, \Rightarrow\,\tilde\varphi(|u|, t,s)\to \frac{\Phi_1(\tilde t,\tilde u)  }{(nu_*^2)^{1/2}}+\frac{\Phi_2(\tilde s, \tilde t,\tilde u) }{nu_*^2}+O(n^{-3/2}),
\end{align}
where $\Phi_1$ is a linear function in $\tilde u,\tilde t$ and $\Phi_2$ is quadratic. Notice that in order to prove Theorem \ref{t:2}, 
we would like to control the terms up to the order $O(n^{-1})$
in (\ref{p1.1}). But since the expansion for $\varphi$ starts from $O(n^{-1/2})$ term, we need
to keep only terms of order $O(1)$ and $O(n^{-1/2})$  in the expansion of other functions   in (\ref{p1.3}).
It is straightforward to see that
\begin{align}\label{exp_h}
nh(|u|)=&nh(u_*)-2c_2\tilde u^2+\tilde c u_*\tilde u^3(nu_*^2)^{-1/2}+O(n^{-1}),\\
-nh(-it)=&-nh(u_*)-2c_2\tilde t^2-i\tilde cu_*\tilde t^3(nu_*^2)^{-1/2}+O(n^{-1}).
\notag\end{align}
Here and below 
\begin{align}\label{c_k}
c_k=n^{-1}\Tr G^{k}(u_*^2)\quad k=2,3
\end{align}
with $G(x)$ defined in (\ref{L,G}), and
 $\tilde c$ is some coefficients whose exact value is not important for us.

Let us  introduce the averaging which is obtained from (\ref{p1.3}), if we make the changes of variables (\ref{ch3})
and use that $n\varepsilon=\tilde\varepsilon/\delta$. 
Using the above asymptotic expansions and keeping only terms of order $O(1)$ and $O(n^{-1/2})$, we obtain
\begin{align}\label{angle}
\left\langle f(\tilde t,\tilde u,\tilde s,\phi,\tilde v)\right\rangle_0&=
\frac{n\cdot e^{-2(u_*/\delta)\tilde\varepsilon}}{2u_*^2}
\int f (\tilde t,\tilde u,\tilde s,\phi,\tilde v)\cdot \frac{e^{-2c_2(\tilde t^2+\tilde u^2)-\tilde s^2}}{(\sqrt \pi)^{3}}
\Big(u_*^2+\frac{u_*(\tilde u-i\tilde t)}{(u_*^2n)^{1/2}}\Big)\\ \notag
&\times J_0\Big(-2i\tilde\varepsilon \big(\frac{u_*}{\delta}+\frac{\tilde u}{\delta(nu_*^2)^{1/2}}\big)\Big )
 \cdot\mathcal V \Big(2\tilde\varepsilon\big(\frac{u_*}{\delta}-\frac{i\tilde t}{\delta(nu_*^2)^{1/2}}\big)\Big)\\
&\times\Big(1+\dfrac{\Phi_*(\tilde t, \tilde u, \tilde s)}{(u_*^2n)^{1/2}}\Big) \Big(\frac{\Phi_1(\tilde t,\tilde u ) }{(u_*^2n)^{1/2}}
+\frac{\Phi_2(\tilde t,\tilde u) }{u_*^2n}\Big)
  d\tilde t d\tilde u d\tilde s.  \notag
 \end{align}
Here the last multiplier at the first line corresponds to $(-it)|u|$ in (\ref{p1.3}) and we  denoted by $\Phi_*(\tilde t, \tilde u, \tilde s)/(u_*^2n)^{1/2}$ 
the sum of all $O(n^{-1/2})$ terms which appear from the exponents of the second line of (\ref{p1.3}). We also set
\begin{align*}
\mathcal{V}(x)=\int e^{-2 x \tilde v^2} \dfrac{d\tilde v}{\sqrt{\tilde v^2+1}},
\end{align*}
and  used that
\begin{align*}
\dfrac{1}{2\pi}\int e^{-2((u_*/\delta)+\frac{\tilde u}{\delta(nu_*^2)^{1/2}})\tilde\varepsilon\cos\varphi}d\phi= 
J_0\Big(-2i\tilde\varepsilon \big((u_*/\delta)+\frac{\tilde u}{\delta(nu_*^2)^{1/2}}\big)\Big ),
\end{align*}
where $J_0$ is a zero-order Bessel function.

Since  for small $\tilde x$ and any fixed $x$ 
\begin{align*}
\dfrac{\int_{-\pi}^{\pi}\cos\phi  \cdot e^{i(x+\tilde x )\cos \phi}  d\phi}{\int_{-\pi}^{\pi} e^{i(x+\tilde x) \cos \phi}  d\phi}=-i\frac{J_0'(x+\tilde x)}{J_0(x+\tilde x)}
 = -i(\log J_0(x))'\Big(1+\frac{(\log J_0(x))''}{(\log J_0(x))'}\tilde x +O(\tilde x^2)\Big),
\end{align*}
using the representation (\ref{p1.4}) we obtain for $|u|=u_*+\frac{\tilde u}{(nu_*^2)^{1/2}}$
\begin{align}\label{angle2}
\frac{\tilde \varepsilon T}{n}=&\left\langle \cos\phi\,|u|\right\rangle+O(n^{-1})=-i\left\langle \frac{J_0'\big(-2i\tilde\varepsilon |u|/\delta\big )}
{J_0\big(-2i\tilde\varepsilon |u|/\delta\big )}|u|\right\rangle
+O(n^{-1})\\
=& -i(\log J_0(-2i\tilde\varepsilon u_*/\delta))'\left\langle\Big(1+\tilde J\frac{\tilde u}{\delta(nu_*^2)^{1/2}}\Big)
\Big(u_*+\frac{\tilde u}{(nu_*^2)^{1/2}}\Big)\right\rangle+O(n^{-1})\notag\\
=& -i(\log J_0(-2i\tilde\varepsilon u_*/\delta))'\Big(u_*\left\langle 1\right\rangle+\Big(\frac{\tilde J u_*}{\delta}+1\Big)\left\langle\frac{\tilde u}{(nu_*^2)^{1/2}}
\right\rangle\Big)+O(n^{-1}),
\notag\end{align}
where 
\[\tilde J=\frac{(\log J_0(x))''}{(\log J_0(x))'}\Big|_{x=-2i\tilde\varepsilon u_*/\delta}.\]
By construction of the averaging (\ref{p1.3}) and (\ref{angle}) we have
\begin{align}\label{angle1}
1=&\mathcal{Z}(\varepsilon,\varepsilon)=\left\langle 1\right\rangle+O(n^{-1}).
\end{align}
Hence, we are left only to compute 
\begin{align*}
\left\langle\frac{\tilde u}{(nu_*^2)^{1/2}}\right\rangle=\left\langle\frac{\tilde u}{(nu_*^2)^{1/2}}\right\rangle_0+O(n^{-1}).
\end{align*}
But it is easy to see that to obtain the term
of the order $O(n^{-1})$, we need to take the term which is linear in $\tilde u n^{-1/2}$ in one of the multipliers of (\ref{angle}).
And if we use some factor different from $\Phi_1$, then $\Phi_1$ will give us additional $n^{-1/2}$ and the result will be
 $O(n^{-3/2})$ or less. Hence
\begin{align} \notag
\left\langle \frac{\tilde u}{(nu_*^2)^{1/2}}\right\rangle_0=&\dfrac12\cdot e^{-2(u_*/\delta)\tilde\varepsilon}\cdot 
J_0( -2i\tilde\varepsilon\frac{u_*}{\delta})\cdot
 \mathcal V(2\tilde\varepsilon \frac{u_*}{\delta})\cdot  \frac{\partial \Phi_1}{\partial \tilde u}\Big|_{\tilde u=0}  
\int   \frac{e^{-2c_2(\tilde t^2+\tilde u^2)-\tilde s^2}}{(\sqrt \pi)^{3}}\tilde u^2 d\tilde t d\tilde u d\tilde s \\ \label{angle_u}
&=\frac{e^{-2(u_*/\delta)\tilde\varepsilon}}{16 c_2^2}\cdot J_0( -2i\tilde\varepsilon\frac{u_*}{\delta})\cdot
 \mathcal V(2\tilde\varepsilon \frac{u_*}{\delta})\cdot \frac{\partial \Phi_1}{\partial \tilde u}\Big|_{\tilde u=0} +O((u_*^2n)^{-1}).
\end{align}
To find $\Phi_1$ we use the asymptotic relations
\begin{align*}
1-n^{-1}\Tr G(-t^2+s^2)=&-\frac{2c_2u_*i\tilde t }{(u_*^2n)^{1/2}}+O((u_*^2n)^{-1}),\\
n^{-1}\Tr G(-t^2+s^2)G(|u|^2)=& c_2-\frac{2c_3u_*(\tilde u-i\tilde t)}{(u_*^2n)^{1/2}}+O((u_*^2n)^{-1})
\end{align*}
with $c_2,c_3$ defined in (\ref{c_k}).

Using the relations in (\ref{p1.2}) we find finally
\begin{align}\label{Phi_1}
\Phi_1=&2c_2^2u_*^3(\tilde u-i\tilde t),
\end{align}
and then  (\ref{angle_u}) implies
\begin{align}\label{angle_u1}
\left\langle \frac{\tilde u}{(nu_*^2)^{1/2}}\right\rangle=&\frac{e^{-2u_*\tilde\varepsilon/\delta}}{8}\cdot  u_*^3\cdot J_0( -2iu_*\tilde\varepsilon/\delta) 
\cdot \mathcal V(2\tilde\varepsilon u_*/\delta)+O((u_*^2n)^{-1}).
\end{align}

\section{Proof of Theorem \ref{t:3}}
Now we perform an asymptotic analysis of $T(z,\varepsilon)$ of (\ref{p1.1_new}) in the intermediate regime between  Theorem \ref{t:1} and Theorem \ref{t:3}: $n^{-1/4}\ll \delta \ll 1$, $\varepsilon\sim (\delta n)^{-1}$. Again, we consider only $\omega\in \Omega_{\epsilon_0/2}^{(1)}\cap \Omega_{\epsilon_0/2}^{(2)}\cap \Omega^{(3)}$ (see conditions (A3) -- (A4)).

Since by (\ref{exp_G}) and (\ref{ti-de_1})
\begin{align*}
n^{-1}\Tr G(u^2)=n^{-1}\Tr G(0)-c_2u^2+O(u^4)=1+k\delta^2-c_2u^2+O(u^4)+O(\delta^4),
\end{align*}
we obtain  that  $u_*$ of (\ref{eq_u*}) has the form
\begin{align}\label{u_*.3}
u_*=\Big(\frac{k}{c_2}\Big)^{1/2}\delta(1+O(\delta^2)).
\end{align}
Doing  the change of variables (\ref{ch1})-(\ref{ch2}) and  choosing the integration contours for $t$ and $R$ (as in
the proof of Theorem \ref{t:2}), we can integrate over $R$ and restrict the integration with respect to $s$ by (\ref{restr_s}).
It is easy to see that under conditions of Theorem \ref{t:3} on $\delta$ we have for $t=iu_*+\tau$, $\tau\ll1$
\begin{align*}
\tilde F(t)=-h(u_*)+\frac{c_2}{2}(2iu_*\tau+\tau^2)^2+O(u_*\tau^3)+O(\tau^4)=-h(u_*)-2c_2u_*^2\tau^2+O(u_*\tau^3)+O(\tau^4).
\end{align*}
Hence, we can restrict the integration with respect to $t$, $s$ and $u$ by (\ref{restr_t1}), (\ref{restr_s}) and (\ref{restr_u}),
and obtain the representation (\ref{p1.3}). 
After the change (\ref{ch3}) the expansion (\ref{exp_h}) takes the form
\begin{align*}
nh(|u|)=& nh(u_*)-2c_2\tilde u^2+O\Big(\frac{\tilde u^3}{n^{1/2}u_*^2}\Big)+O\Big(\frac{\tilde u^4}{nu_*^4}\Big),\\
-nh(it) =&-nh(u_*)-2c_2\tilde\tau^2+O\Big(\frac{\tilde\tau^3}{n^{1/2}u_*^2}\Big)+O\Big(\frac{\tilde\tau^4}{nu_*^4}\Big).
\end{align*}
Notice, that if $\delta\sim n^{-1/4},\, u_*\sim n^{-1/4}$ (like in Theorem \ref{t:1}), then the second and the third terms of the expansion
are of the same order as the first one, and we cannot expand with respect  to them. But for  $\delta\gg n^{-1/4}$, these terms are small
and we can write expansions similar to that from the proof of Theorem \ref{t:2}, but the remainder terms are of the order
$O\big((nu_*^4)^{-1}\big)$

Repeating the argument used in the proof of Theorem \ref{t:2}, we introduce the averaging (\ref{p1.3}) and use (\ref{angle2}) in the form
\begin{align*}
\frac{\tilde \varepsilon T}{n}=& -i(\log J_0(-2i\tilde\varepsilon u_*/\delta))'\Big(u_*\left\langle 1\right\rangle+\big(\frac{\tilde J u_*}{\delta}+1\big)
\left\langle\frac{\tilde u}{(nu_*^2)^{1/2}}\right\rangle\Big)+O\big((\delta (u_*^2 n)^{1/2})^{-1}\big).
\notag\end{align*}
Then (\ref{angle1}) implies that  we are left to find $\left\langle\tilde u/(nu_*^2)^{1/2}\right\rangle$. This can be done if  we expand 
all multipliers under the integral in the r.h.s  of (\ref{p1.3}) and keep the main term in these expansion. 
Using the more advanced form of (\ref{exp_phi_til})
\begin{align*}
\tilde\varphi(|u|, t,s)=
 \frac{u_*^3\tilde\Phi_1(\tilde t,\tilde u)  }{(nu_*^2)^{1/2}}+\frac{u_*^2\tilde\Phi_2(\tilde s, \tilde t,\tilde u) }{nu_*^2}
+\frac{u_*\tilde\Phi_3(\tilde t,\tilde u,s)  }{(nu_*^2)^{3/2}}+O((nu_*^2)^{-2}),
\end{align*}
where $\tilde\Phi_{1,2,3}$ are bounded for bounded $\tilde t,\tilde u,\tilde s$, one can check that for $u_*\sim\delta\gg n^{-1/4}$ the main term of $\tilde\varphi(|u|, t,s)$ is the first one in the above expansion.
Then we get (\ref{angle_u}) in the form
\begin{align}\label{angle_u2}
\left\langle \frac{\tilde u}{(nu_*^2)^{1/2}}\right\rangle_0=&\frac{e^{-2(u_*/\delta)\tilde\varepsilon}}{16 c_2^2}\cdot J_0( -2i\tilde\varepsilon\frac{u_*}{\delta})\cdot
 \mathcal V(2\tilde\varepsilon \frac{u_*}{\delta})\cdot u_*^3\frac{\partial\tilde \Phi_1}{\partial \tilde u}\Big|_{\tilde u=0} \Big(1+O((u_*^4n)^{-1})\Big).
\end{align}
Since it follows from (\ref{Phi_1}) that
\begin{align*}
\tilde\Phi_1=&2c_2^2(\tilde u-i\tilde t),
\end{align*}
we obtain (\ref{angle_u1}) with an error multiplier  $1+O((u_*^4n)^{-1})$.


\section{Appendix}

\subsection{Grassmann integration}
Let us consider two sets of formal variables
$\{\psi_j\}_{j=1}^n,\{\overline{\psi}_j\}_{j=1}^n$, which satisfy the anticommutation
conditions
\begin{equation}\label{anticom}
\psi_j\psi_k+\psi_k\psi_j=\overline{\psi}_j\psi_k+\psi_k\overline{\psi}_j=\overline{\psi}_j\overline{\psi}_k+
\overline{\psi}_k\overline{\psi}_j=0,\quad j,k=1,\ldots,n.
\end{equation}
Note that this definition implies $\psi_j^2=\overline{\psi}_j^2=0$.
These two sets of variables $\{\psi_j\}_{j=1}^n$ and $\{\overline{\psi}_j\}_{j=1}^n$ generate the Grassmann
algebra $\mathfrak{A}$. Taking into account that $\psi_j^2=0$, we have that all elements of $\mathfrak{A}$
are polynomials of $\{\psi_j\}_{j=1}^n$ and $\{\overline{\psi}_j\}_{j=1}^n$ of degree at most one
in each variable. We can also define functions of
the Grassmann variables. Let $\chi$ be an element of $\mathfrak{A}$, i.e.
\begin{equation}\label{chi}
\chi=a+\sum\limits_{j=1}^n (a_j\psi_j+ b_j\overline{\psi}_j)+\sum\limits_{j\ne k}
(a_{j,k}\psi_j\psi_k+
b_{j,k}\psi_j\overline{\psi}_k+
c_{j,k}\overline{\psi}_j\overline{\psi}_k)+\ldots.
\end{equation}
For any
sufficiently smooth function $f$ we define by $f(\chi)$ the element of $\mathfrak{A}$ obtained by substituting $\chi-a$
in the Taylor series of $f$ at the point $a$. Since $\chi$ is a polynomial of $\{\psi_j\}_{j=1}^n$,
$\{\overline{\psi}_j\}_{j=1}^n$ of the form (\ref{chi}), according to (\ref{anticom}) there exists such
$l$ that $(\chi-a)^l=0$, and hence the series terminates after a finite number of terms and so $f(\chi)\in \mathfrak{A}$.

Following Berezin \cite{Ber}, we define the operation of
integration with respect to the anticommuting variables in a formal
way:
\begin{equation*}
\intd d\,\psi_j=\intd d\,\overline{\psi}_j=0,\quad \intd
\psi_jd\,\psi_j=\intd \overline{\psi}_jd\,\overline{\psi}_j=1,
\end{equation*}
and then extend the definition to the general element of $\mathfrak{A}$ by
the linearity. A multiple integral is defined to be a repeated
integral. Assume also that the ``differentials'' $d\,\psi_j$ and
$d\,\overline{\psi}_k$ anticommute with each other and with the
variables $\psi_j$ and $\overline{\psi}_k$. Thus, according to the definition, if
$$
f(\psi_1,\ldots,\psi_k)=p_0+\sum\limits_{j_1=1}^k
p_{j_1}\psi_{j_1}+\sum\limits_{j_1<j_2}p_{j_1,j_2}\psi_{j_1}\psi_{j_2}+
\ldots+p_{1,2,\ldots,k}\psi_1\ldots\psi_k,
$$
then
\begin{equation*}
\intd f(\psi_1,\ldots,\psi_k)d\,\psi_k\ldots d\,\psi_1=p_{1,2,\ldots,k}.
\end{equation*}

   Let $A$ be an ordinary Hermitian matrix with a positive real part. The following Gaussian
integral is well-known
\begin{equation}\label{G_C}
\intd \exp\Big\{-\sum\limits_{j,k=1}^nA_{jk}z_j\overline{z}_k\Big\} \prod\limits_{j=1}^n\dfrac{d\,\Re
z_jd\,\Im z_j}{\pi}=\dfrac{1}{\mdet A}.
\end{equation}
One of the important formulas of the Grassmann variables theory is the analog of this formula for the
Grassmann algebra (see \cite{Ber}):
\begin{equation}\label{G_Gr}
\int \exp\Big\{-\sum\limits_{j,k=1}^nA_{jk}\overline{\psi}_j\psi_k\Big\}
\prod\limits_{j=1}^nd\,\overline{\psi}_j d\,\psi_j=\mdet A,
\end{equation}
where $A$ now is any $n\times n$ matrix.

Let
\[
F=\left(\begin{array}{cc}
A&\chi\\
\eta&B\\
\end{array}\right),\quad
\]
where $A$ and $B$ are Hermitian complex $k\times k$ matrices such that $\Re B>0$ and $\chi$, $\eta$ are $k\times k$ matrices
of independent
anticommuting Grassmann variables, and let
\[
\Phi=(\psi_{1},\ldots,\psi_{k},x_{1},\ldots,x_{k})^t,
\]
where $\{\psi_j\}_{j=1}^k$ are independent Grassmann variables and $\{x_j\}_{j=1}^k$ are complex variables.
Combining (\ref{G_C}) -- (\ref{G_Gr}) we obtain (see \cite{Ber})
\begin{equation}\label{G_comb}
\intd \exp\{-\Phi^+F\Phi\}\prod\limits_{j=1}^kd\overline{\psi}_j\, d\psi_j \prod\limits_{j=1}^k\dfrac{\Re x_j\Im
x_j}{\pi}=\Sdet^{-1}\, F,
\end{equation}
where
\begin{equation}\label{sdet}
 \Sdet\, F=\dfrac{\det\,(B-\eta\, A^{-1}\,\chi)}{\det\, A}.
\end{equation}
Notice also that if we define
\begin{equation}\label{Str}
\Str\,F=\Tr B -\Tr A, 
\end{equation}
then
\begin{equation}
\log \Big(\Sdet F\Big)=\Str\Big(\log F\Big).
\end{equation}

We will need also the following Hubbard-Stratonovich transform formulas based on Gaussian integration.
\begin{align}\label{Hub_C}
&e^{ab}=\pi^{-1} \int e^{a \bar u+b u-\bar u u} d\bar u\, du,\\ \label{Hub_Gr}
&e^{-\rho\tau}=\int e^{\rho \chi+\tau\eta+\chi\eta} d\eta d\chi.
\end{align}
Here $a, b$ can be complex numbers or sums of the products of even numbers of Grassmann variables (i.e. commuting elements of Grassmann algebra),
and $\rho,\tau$ are sums of the products of odd numbers of Grassmann variables (i.e. anticommuting elements of Grassmann algebra).


\end{document}